\documentclass[preprint,pre,aps,showpacs]{revtex4}
\usepackage{graphicx}
\begin{document}

\title{Comparison of Statistical Multifragmentation Model simulations
with Canonical Thermodynamic Model results: a few representative cases} 

\author{A. Botvina$^{1,2}$, G. Chaudhuri$^3$, S. Das Gupta$^3$, and 
I. Mishustin$^{2,4}$}

\affiliation{$^1$Institute for Nuclear
Research, Russian Academy of Sciences, 117312 Moscow, Russia} 

\affiliation{$^2$Frankfurt Institute for Advanced Studies, J.W. Goethe
University, D-60438 Frankfurt am Main, Germany} 

\affiliation{$^3$Physics Department, McGill University, 
Montr{\'e}al, Canada H3A 2T8}

\affiliation{$^4$Kurchatov Institute, Russian Research Center,
123182 Moscow, Russia}

\date{\today}

\begin{abstract}

The statistical multifragmentation model (SMM) has been widely used
to explain experimental data of intermediate energy heavy ion collisions.
A later entrant in the field is the canonical thermodynamic model (CTM) 
which is also being used to fit experimental data.  The basic physics
of both the models is the same, namely that fragments are produced
according to their statistical weights in the available phase space.  
However, they are based on different statistical ensembles, and the 
methods of calculation are different: while the SMM uses 
Monte-Carlo simulations, the CTM solves recursion relations.  
In this paper we compare the predictions of the two models for a few 
representative cases.

\end{abstract}


\maketitle

\section{Introduction}

The statistical multifragmentation model (SMM) has had wide applications 
in the field of intermediate energy heavy ion collisions.  The basic
assumption is that when two heavy ions collide, after pre-equilibrium 
emission, they form a compound system 
with given mass number $A_0$, charge number $Z_0$ and given energy
$E_0$.  This system will expand and break up into different composites.
In this expanded volume the nuclear interaction between different
composites can be neglected and Coulomb interaction can be included in 
an approximate way.  Assuming that the populations of different channels
are solely given by statistical weights in the available phase space, the 
average yield of given species with a neutron number $N$ and proton number 
$Z$ can be computed by Monte-Carlo simulation.  Details may be found in 
\cite{Bondorf1}.
Many versions of the Monte-Carlo simulations are possible.  For example,
one can choose the freeze-out volume to be dependent on the multiplicity
of a channel. For reactions where equilibrated sources are formed, 
the SMM gives very good description of experimental data 
\cite{Bondorf1,Xi,Dagostino,Scharenberg}.

The canonical thermodynamic model (CTM) characterises the disintegrating
system by a temperature $T$, the mass number $A_0$, charge number $Z_0$
and a fixed freeze-out volume.  With these simplifications, computations
of average yields of composites are straightforward and do not need
Monte-Carlo simulations.  Details can be found in \cite{Das1}.  This
model has also been used successfully to fit experimental data
\cite{Das1,Tsang1}.  It is expected that results will be similar to those
obtained from the SMM.  Here we embark on more quantitative comparison
of the results generated in the two frameworks for a few representative
cases.  Such a comparison is long overdue.

\section{Parameters of the CTM and SMM used in the present work}

First, we give description of the CTM.  For the SMM 
more than one approach was used which will be described below.
Details of the computational procedures for the canonical model can
be found elsewhere \cite{Das1,Chaudhuri1} hence we provide here only
the bare minimal.

The fragmenting system has $N_0$ neutrons and $Z_0$ neutrons. 
We consider all break-up channels (partitions) $\{p\}$, and define 
$n_{i,j}$ as the number of fragments with neutron number $i$ and 
the number of protons $j$. As is well known, the number of partitions 
of medium and heavy systems $(N_0\sim Z_0\sim 100)$ is enormous 
(see e.g. \cite{Jackson}). 
The canonical partition function for non-interacting fragments is given by 
\begin{eqnarray}
Q_{N_0,Z_0}=\sum_{\{p\}}\prod_{i,j} 
\frac{\omega_{i,j}^{n_{i,j}}}{n_{i,j}!}
\end{eqnarray}
Here the sum is over all possible partitions, which satisfy 
the conditions $N_0=\sum i\cdot n_{i,j}$ and $Z_0=\sum j\cdot n_{i,j}$; 
$\omega_{i,j}$ is the partition function of one composite with
neutron number $i$ and proton number $j$, respectively. 
The one-body partition
function $\omega_{i,j}$ is a product of two parts: one arising from
the translational motion of the composite and another from the
intrinsic partition function of the composite:
\begin{eqnarray}
\omega_{i,j}=\frac{V_f}{h^3}(2\pi m_{N}aT)^{3/2}\cdot z_{i,j}(int)
\end{eqnarray}
where $m_N\approx 939$ MeV is the average nucleon mass, and 
$a=i+j$ is the mass number of the composite ($i,j$). Here 
$V_f$ is the volume available for translational motion; $V_f$ will
be less than $V$, the volume to which the system has expanded at
break up. We use $V_f = V - V_0$ , where $V_0$ is the normal volume of  
nucleus with $Z_0$ protons and $N_0$ neutrons.

The probability of a given channel $P(\vec n_{i,j})\equiv P(n_{0,1},
n_{1,0},n_{1,1}......n_{i,j}.......)$ is given by
\begin{eqnarray}
P(\vec n_{i,j})=\frac{1}{Q_{N_0,Z_0}}\prod_{i,j}
\frac{\omega_{i,j}^{n_{i,j}}}{n_{i,j}!}
\end{eqnarray}
The average number of composites with $i$ neutrons and $j$ protons is
seen easily from the above equation to be
\begin{eqnarray}
\langle n_{i,j}\rangle=\omega_{i,j}\frac{Q_{N_0-i,Z_0-j}}{Q_{N_0,Z_0}}
\end{eqnarray}
The constraints $N_0=\sum i\cdot n_{i,j}$ and $Z_0=\sum j\cdot n_{i,j}$
can be used to obtain different looking but equivalent recursion relations
for partition functions.  For example
\begin{eqnarray}
Q_{N_0,Z_0}=\frac{1}{N_0}\sum_{i,j}i\omega_{i,j}Q_{N_0-i,Z_0-j}
\end{eqnarray}
These recursion relations allow one to calculate $Q_{N_0,Z_0}$

We list now the properties of the composites used in this work.  The
proton and the neutron are fundamental building blocks 
thus $z_{1,0}(int)=z_{0,1}(int)=2$ 
where 2 takes care of the spin degeneracy.  For
deuteron, triton, $^3$He and $^4$He we use $z_{i,j}(int)=(2s_{i,j}+1)\exp(-
\beta e_{i,j}(gr))$ where $\beta=1/T$, $e_{i,j}(gr)$ is the ground state energy
of the composite and $(2s_{i,j}+1)$ is the experimental spin degeneracy
of the ground state.  Excited states for these very low mass
nuclei are not included.  
For mass number $a=5$ and greater we use
the liquid-drop formula.  For nuclei in isolation, this reads 
\begin{eqnarray}
z_{i,j}(int)=\exp[-\frac{F_{i,j}}{T}]
\end{eqnarray}
Here $F_{i,j}$ is the internal free energy of species $(i,j)$:
\begin{eqnarray}
F_{i,j}=-W_0a+\sigma(T)a^{2/3}+\kappa\frac{j^2}{a^{1/3}}
+s\frac{(i-j)^2}{a}-\frac{T^2a}{\epsilon_0}. 
\end{eqnarray}
The expression includes the 
volume energy, the temperature dependent surface energy, the Coulomb
energy and the symmetry energy.  The term $\frac{T^2a}{\epsilon_0}$
represents contribution from excited states
since the composites are at a non-zero temperature. For nuclei with 
$A$=5 we include
$Z$=2 and 3 and for $A$=6 we include $Z$=2,3 and 4.  For higher masses
we compute the drip lines using the liquid-drop formula above and
include all isotopes within these boundaries.

In the grand canonical formulation for the SMM \cite{Bondorf1}, 
after integrating out translational degrees of freedom, one can 
write the mean multiplicity of nuclear fragments with $i$ and $j$ as 
\begin{eqnarray}
\label{naz} \langle n_{i,j} \rangle =
(2s_{i,j}+1)\frac{V_{f}}{\lambda_T^3}a^{3/2} {\rm
exp}\left[-\frac{1}{T}\left(F_{i,j}(T,V)-\mu a-\nu
j\right)\right]. \nonumber
\end{eqnarray}
Here $\lambda_T=\left(2\pi\hbar^2/m_NT\right)^{1/2}$ is the
nucleon thermal wavelength. 
The chemical potentials $\mu$ 
and $\nu$ are found from the mass and charge constraints:
\begin{equation} \label{eq:ma2}
\sum_{i,j}\langle n_{i,j}\rangle a=A_{0},~~ \sum_{i,j}\langle
n_{i,j}\rangle j=Z_{0}.
\end{equation}
We use the same treatment of the lightest particles with $A<5$ as in 
the canonical case, i.e., the free energy is just their binding energy. 
However, for large fragments we take into account all possible 
charges starting from $Z=2$ up to $Z=A-2$. We have checked that this extension 
beyond the drip line has a very small influence on final results in 
the range of nuclei covered by experiments.  The grand canonical
occupations are used for Monte-Carlo fragment generation in the SMM.

At small excitation energies the standard SMM code \cite{Bondorf1} uses a 
microcanonical treatment, however, taking into account a limited 
number of disintegration channels: as a rule, only partitions with total 
fragment multiplicity 
$M \leq 3$ are considered. This is a very reasonable approximation at 
low temperature, when the compound nucleus and low-multiplicity channels 
dominate. Recently, a full microcanonical version of the SMM using 
the Markov Chain method was introduced \cite{Jackson,Botvina01}. It can 
be used for exploring all partitions without limitation. 
However, it is a more time consuming approach, and it is used in special 
cases only \cite{Botvina01}.

Within the microcanonical ensemble the 
statistical weight of a partition $p$ is calculated as
\begin{eqnarray}
W_{\rm p} \propto exp~S_{\rm p},
\end{eqnarray}
where $S_{\rm p}$ is the corresponding entropy, which depends on fragments 
in this partition, as well as on the excitation energy $E_0$, mass 
number $A_{0}$, charge $Z_{0}$ and volume $V$ of the system. In the 
present work we follow a description which corresponds to approximate 
microcanonical ensemble. Namely, we introduce a temperature $T_{\rm p}$ 
characterising all final states in each partition $p$. It is determined 
from the energy balance equation taking into account the total excitation 
energy $E_0$ \cite{Bondorf1}. 
In the following we determine $S_{\rm p}$ for the found $T_{\rm p}$ by using 
conventional thermodynamical relations. For instance, in this work, 
it can be written as 
\begin{eqnarray}
S_{\rm p}=ln(\prod_{i,j}(2s_{i,j}+1))+ln(\prod_{i,j}a^{3/2})
-ln(A_0^{3/2})-ln(\prod_{i,j}n_{i,j}!)+
\nonumber
\\
(M-1)ln(V_f/\lambda_{T_{p}}^3)
+1.5(M-1)+\sum_{i,j}(\frac{2T_{p}a}{\epsilon_0}-
\frac{\partial \sigma(T_{p})}{\partial T_{p}}a^{2/3}) ,
\end{eqnarray}
where summations are performed over all fragments of the partition $p$. 
We enumerate all considered partitions and select one of them according 
to its statistical weight by the Monte-Carlo method.

At high excitation energy the standard SMM code makes a transition to 
the grand-canonical ensemble, and generate partitions by Monte-Carlo 
sampling from the grand-canonical distribution \cite{Bondorf1}. 
In all ensembles the long range Coulomb interaction between
different composites is included in the Wigner-Seitz approximation.  
For details of this incorporation see references \cite{Bondorf1,Das1}.
In all calculations later on we use the fixed break-up volume 
$V=6V_0$ independently of fragment multiplicity.

\section{Fragment mass distributions}

One major goal of this work is to compare the average yields 
$\langle n_{i,j} \rangle$ of a composite with neutron number
$i$ and proton number $j$, when systems with 
different temperatures dissociate.  In the canonical model this 
is simply given by eq.(4).  For brevity we will display most often
$\langle n_a \rangle\equiv \sum_{a=i+j}\langle n_{i,j} \rangle$.
While for the CTM one needs to specify which
composites are allowed in channels, this is not needed for
SMM simulations.  
We consider a medium-mass nucleus with $A_0$=73 and 
$Z_0$=32 at low excitation energy.  Such a compound system can be 
formed by $^{64}$Ni projectile on $^9$Be and has actually been studied 
experimentally \cite{Mocko1}. In order to demonstrate the mass effect 
on the fragmentation picture we take also two-times larger sources.

First, we start from low excitation energies, and consider 
$E_0 = 1$ MeV/nucleon to perform calculations with 
the microcanonical version of the SMM described above.  
At such low excitation energy the multiplicity should 
be low and for simplicity we restrict the microcanonical
ensemble to multiplicities $M=$ 1, 2 and 3.  
The average $T_{\rm p}$ (see section II) for the excitation energy of 
1 MeV/nucleon is 3.13 MeV. We perform the CTM calculations 
for this temperature.  We will get an exact answer for 
the canonical model, but only an approximate one for the SMM because of 
restriction of multiplicity. Another reason for disagreement between 
the two models is that in the SMM some channels, even at low 
multiplicity, will be just prohibited because of high Coulomb barrier. 
We can therefore hope that when the yield $\langle n_{i,j} \rangle$ is 
significant the agreement between the two models can emerge but for very 
small yields there will be deviations. 
Fig. 1 shows that for important yields, the agreement between
the two models is very good.  This also shows that, at least in the
present example, the average yields are about the same whether 
excitation energy per particle is kept constant or an appropriate
temperature is kept fixed. In real physical processes, as we know, 
the compound nucleus channel ($M=1$) dominates at low excitations. 
Both models describe correctly a transition to this low-energy limit. 

Now we turn to medium and high excitation energies, which lead 
to real multifragmentation. In Fig.~2 we compare the CTM and SMM models  
at intermediate temperatures of 5 and 8 MeV for three systems. 
One system is the same as before ($A_0=73, Z_0=32$), another system
is twice of this size ($A_0=146, Z_0=64$), and the third one is more 
neutron rich ($A_0=146, Z_0=56$). The canonical model calculations are 
the same as before, but the SMM simulations are dictated by convenience 
and simplicity, and are performed with the grand canonical version, 
as discussed.  Instead of excitation energy per nucleon, 
we consider fixed temperature and the starting point of Monte-Carlo
simulations is grand canonical population of composites. 
Remarkable agreement is obtained between the two models
except when $\langle n_a \rangle$ is very small.  We will comment
upon the discrepancies for very small populations shortly.
However, as is clear from the comparison, both models can be 
used in multifragmentation region.

\section{Isotope yields}

Further important details about characteristics of produced fragments 
are given in Figs.~3 and 4.  For selected isotopes we compare the yields 
in the two models as the neutron number changes. Now we are directly 
comparing individual values of $\langle n_{i,j} \rangle$ rather 
than the sum of all isobars. We see once again, unless the multiplicities 
are really small, the yields are very close. Given the choice of composites 
allowed in any channel, the CTM results are exact. In the grand 
canonical SMM calculations we get broader isotope distributions, 
especially for big fragments. This feature explains 
the increased yield of big fragments seen in SMM results in Fig.~2.

The reason is that the grand-canonical ensemble increases the phase 
space, since it takes into account many more partitions which 
include exotic nuclei, and conserves the mass number and charge 
conservation laws on the average only. Since the number of exotic nuclei 
is greater for large nuclei we obtain more deviation at large A and Z. 
We can make a quite 
general statement: any restriction of the partition phase space
lead to more narrow isotope distributions. We expect narrow distributions 
also for the microcanonical ensemble, since it imposes the exact energy 
conservation in the partitions additionally.

\section{Caloric curves}

The caloric curves, temperature $T$ against excitation energy $E^*/A_0$ 
agree very well in the two models. This dependence is very important for 
statistical systems, and it 
can be in many cases extracted experimentally via calorimetry. Examples of 
the caloric curves are shown in Fig.5. We conclude that 
small differences in fragments yields do not 
influence global thermodynamical characteristics of the system. 

We have calculated the caloric curves for all three sources under 
consideration, and they look practically identical. The experimental 
identification of the difference $\Delta T \sim 0.1$ MeV is probably 
impossible. However, we would like to draw attention to the interesting 
mass and isospin effects which can be seen by comparing different sources, 
and which may be important in big and neutron rich astrophysical systems. 
This can be seen in fig.6 showing differences in excitation 
energies $E^*/A_0(A_0=146, Z_0=64)-E^*/A_0(A_0=146, Z_0=56)$ and 
$E^*/A_0(A_0=146, Z_0=64)-E^*/A_0(A_0=73, Z_0=32)$ versus temperature. 
In the region of multifragmentation ($T > 4$ MeV) a bigger system has 
a slightly larger excitation energy per nucleon. This is a result of 
stronger Coulomb effect 
in big systems. An isospin effect can also be seen at the transition from the 
channels with big (compound-like) fragments to the full multifragmentation 
region, i.e., at temperatures $T \approx 3.5-5$ MeV: The neutron rich 
system has a slightly lower excitation energy.

\section {Summary}

We have compared the yields of composites, as well as the caloric curves, 
in two different model prescriptions corresponding to different ensembles. 
Generally they agree well.  The SMM is more versatile: one can use either 
the excitation energy or the temperature as the fixed parameter. It also
allows for other modifications such as multiplicity dependence
of the freeze-out volume.  The disadvantage is that it is computationally
quite involved.  The canonical model is more restrictive but extremely
easy to use. We have found some differences in results of different 
models, in particular, concerning isotope distributions. This may cause 
different interpretation of some experimental observables, e.g., the 
isoscaling characteristics. 
We should remark that in this paper we analyse the yields of 
primary hot fragments. The subsequent de-excitation of these excited 
fragments must be taken into account in interpretation of experimental 
data \cite{Bondorf1,Das1}. This may lead to additional deviations between 
predictions of microcanonical and canonical/grandcanonical ensembles, 
since the microcanonical SMM treatment allows for temperature fluctuations 
in different partitions.

In this respect, the important question is what statistical ensemble 
would be preferable for analysis of experimental characteristics sensitive 
to the kind of ensemble. The answer can be provided by experimental 
information only. If the data have 
clear evidences of production of a thermal source with well determined 
$A_0$, $Z_0$ and $E_0$ the microcanonical ensemble is justified. If one can 
speak only about exact $A_0$ and $Z_0$ conservations, the canonical
ensemble is the 
best. If experiment can give only average values for parameters of the 
thermal source, the grand canonical model can be applied. 
Moreover, as is well known, the preequilibrium dynamical stage can 
provide a very broad distribution in parameters of the thermal sources. 
Including all this distribution may have a greater impact on 
the results, than selection of a particular ensemble. 
In the case of an experimental uncertainty with 
parameters of the thermal source, we believe, the evaluation of 
experimental data should include fitting the data within one of the 
statistical models, and addressing physical characteristics 
which are universal for all ensembles, like break-up volume and internal 
properties of hot primary fragments.

\begin{figure}
\includegraphics[width=5.5in,height=4.5in,clip]{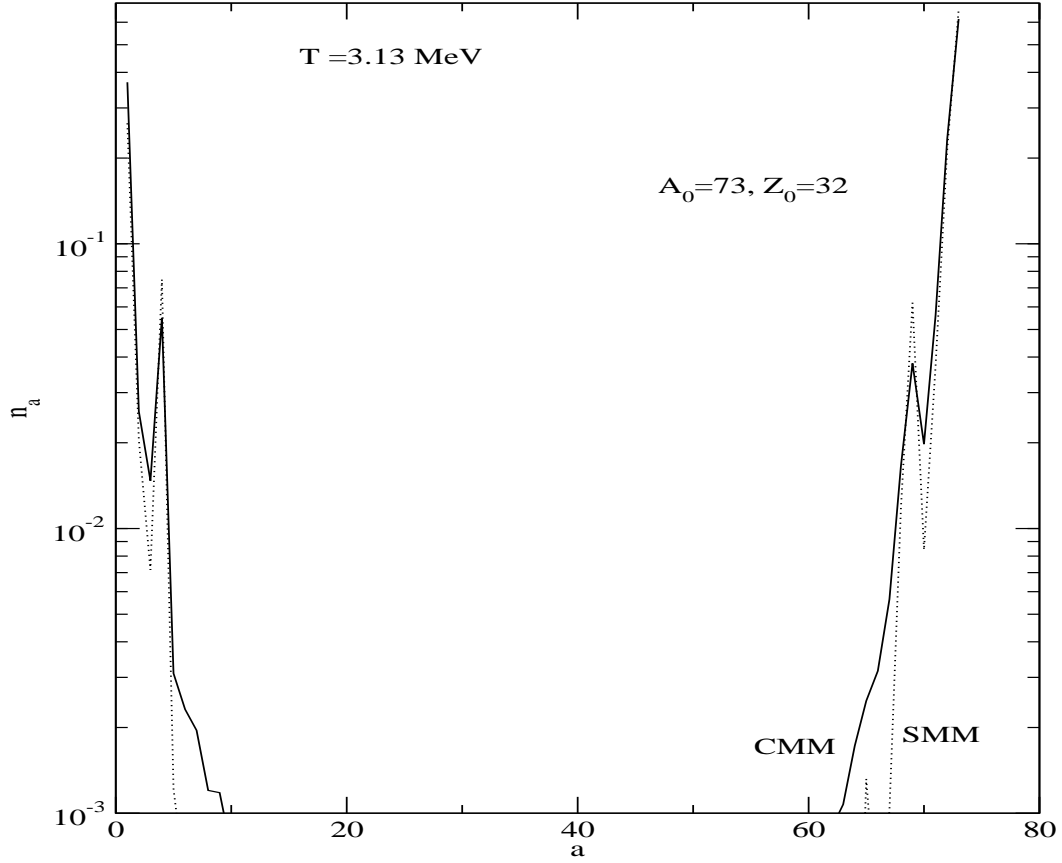}
\caption{ 
Mass distributions of fragments as predicted by the CTM (solid lines) 
and the SMM (dashed lines). The SMM calculations are for restricted 
microcanonical ensemble with excitation energy of 1 MeV/nucleon, 
corresponding to the temperature of 3.13 MeV taken as input in the CTM. 
The dissociating system is $A_0=73, Z_0=32$.}
\end{figure}

\begin{figure}
\includegraphics[width=5.5in,height=4.5in,clip]{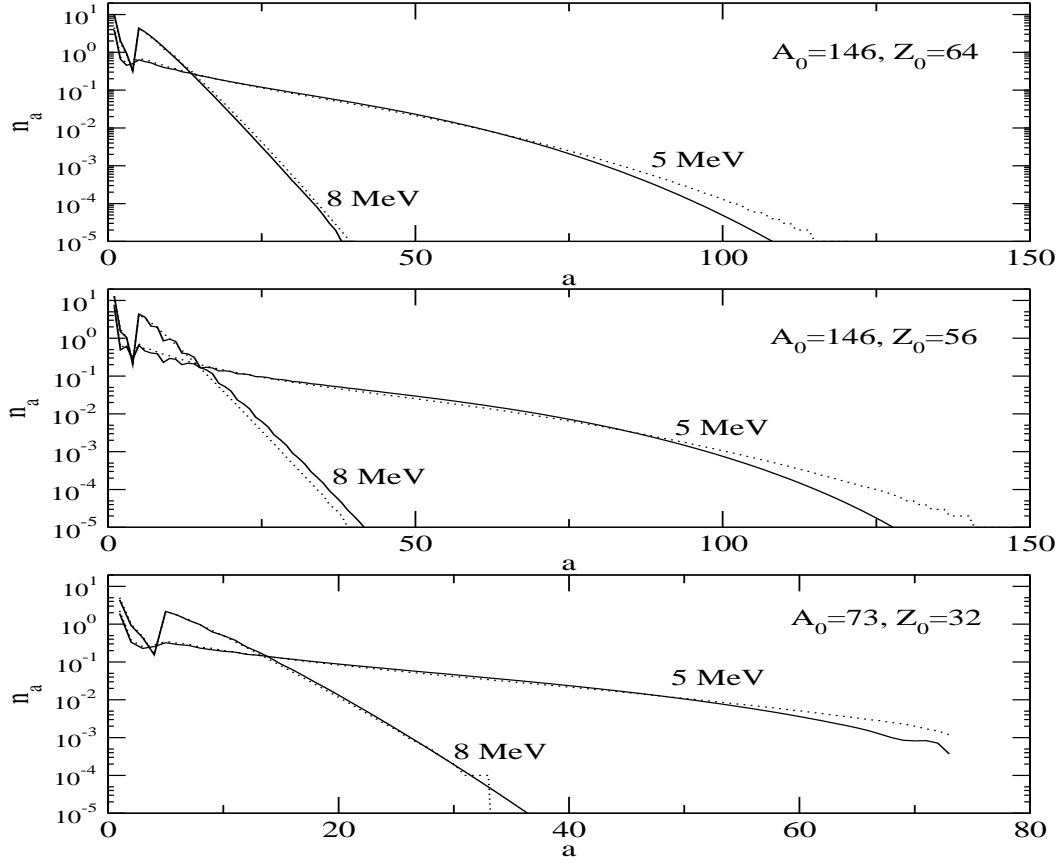}
\caption{ 
Fragment mass distributions $n_a$ predicted by the CTM (solid lines) and the 
SMM (dashed lines) at temperatures 5 and 8 MeV, for three dissociating 
systems indicated in the figure. The SMM calculations are based on the 
grand canonical ensemble.}
\end{figure}

\begin{figure}
\includegraphics[width=5.5in,height=4.5in,clip]{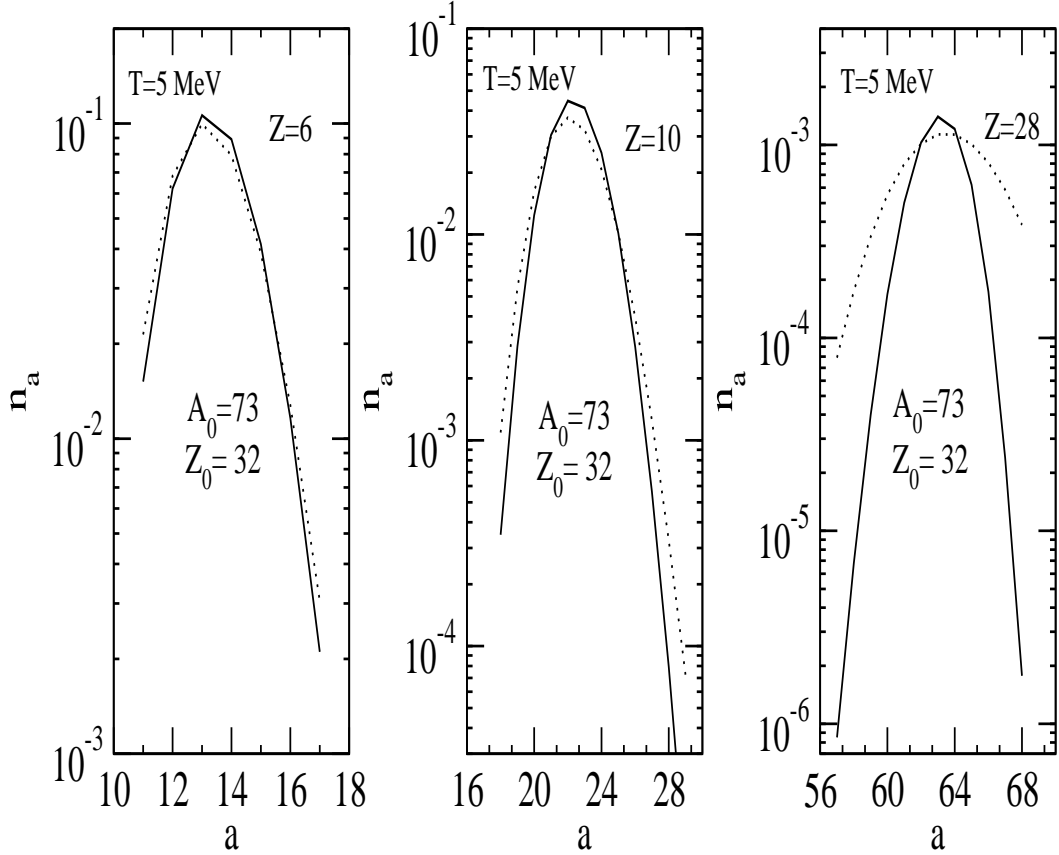}
\caption{ 
Yields of individual isotopes (indicated in the figure) as 
predicted by the CTM (solid lines) and the SMM (dotted lines). 
Results are presented for 
the dissociating system $A_0=73, Z_0=32$ and temperature
5 MeV.}
\end{figure}

\begin{figure}
\includegraphics[width=5.5in,height=4.5in,clip]{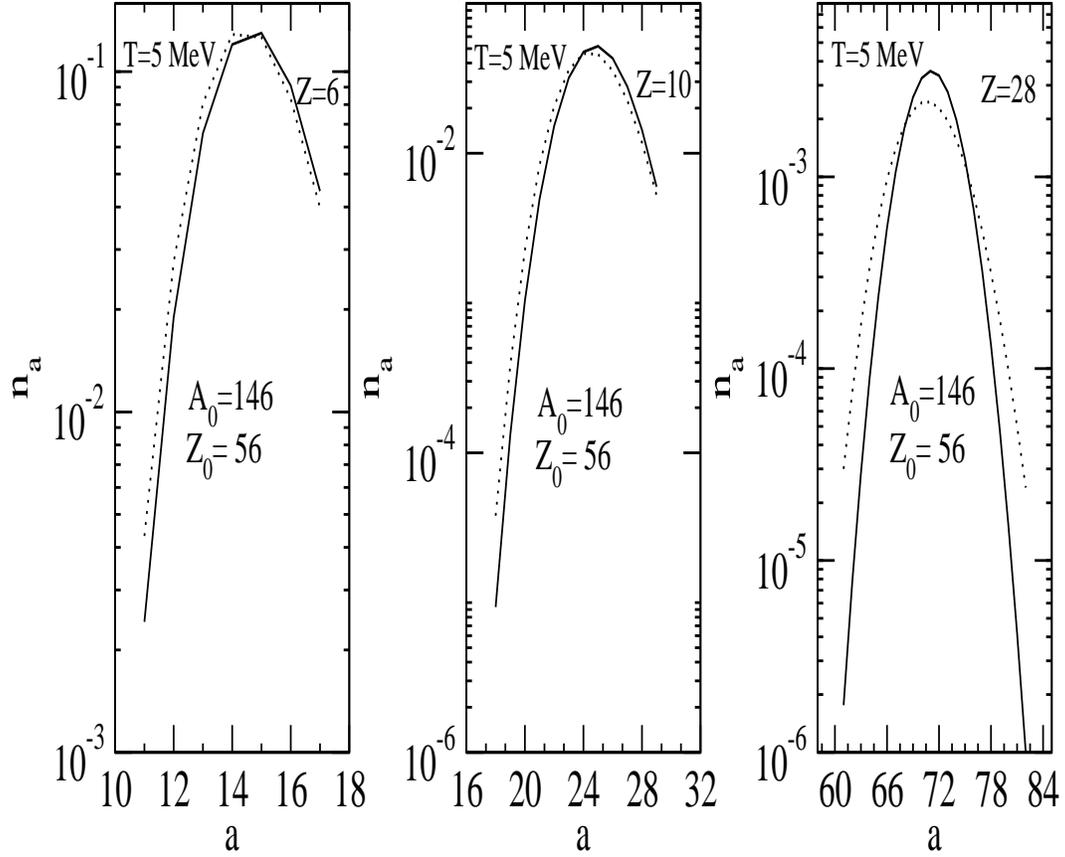}
\caption{ The same as in Fig.~3 except that the dissociating system
is $A_0=146, Z_0=56$.}
\end{figure}

\begin{figure}
\includegraphics[width=5.5in,height=4.5in,clip]{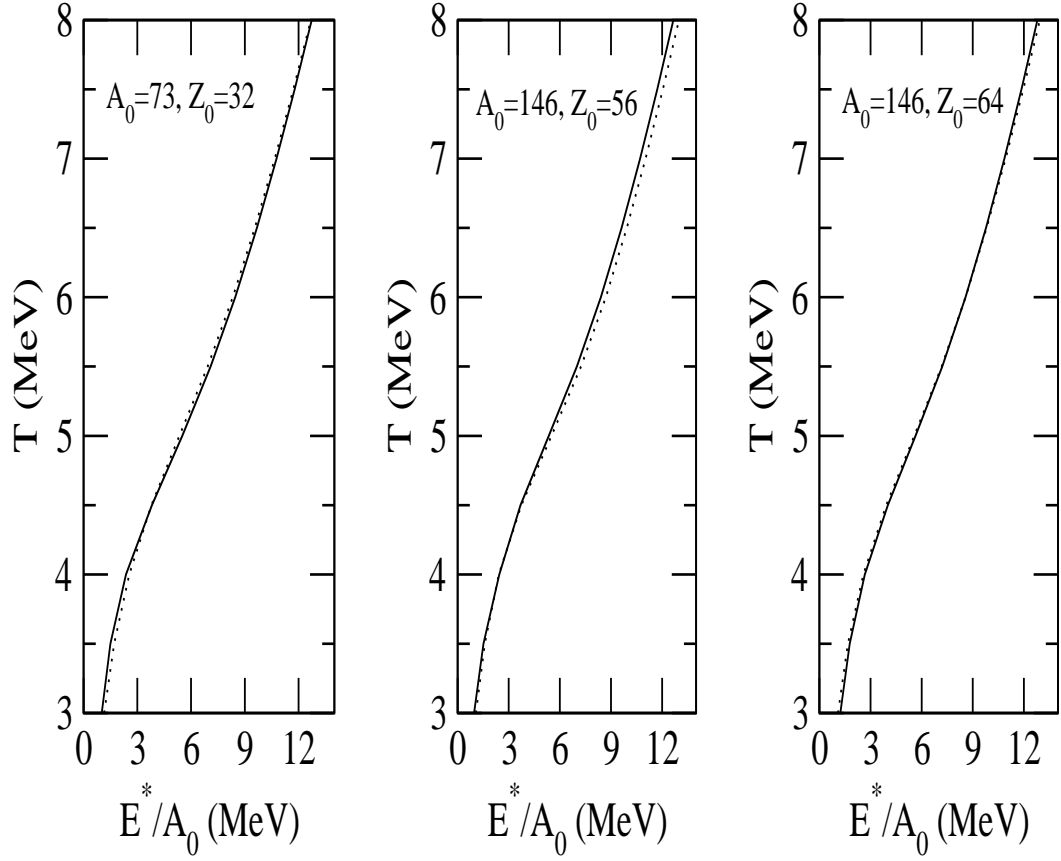}
\caption{The caloric curves ($T$ against $E*/A_0$) predicted 
by the CTM (solid line) and SMM (dotted line) for three 
dissociating systems.}
\end{figure}

\begin{figure}
\includegraphics[width=5.5in,height=4.5in,clip]{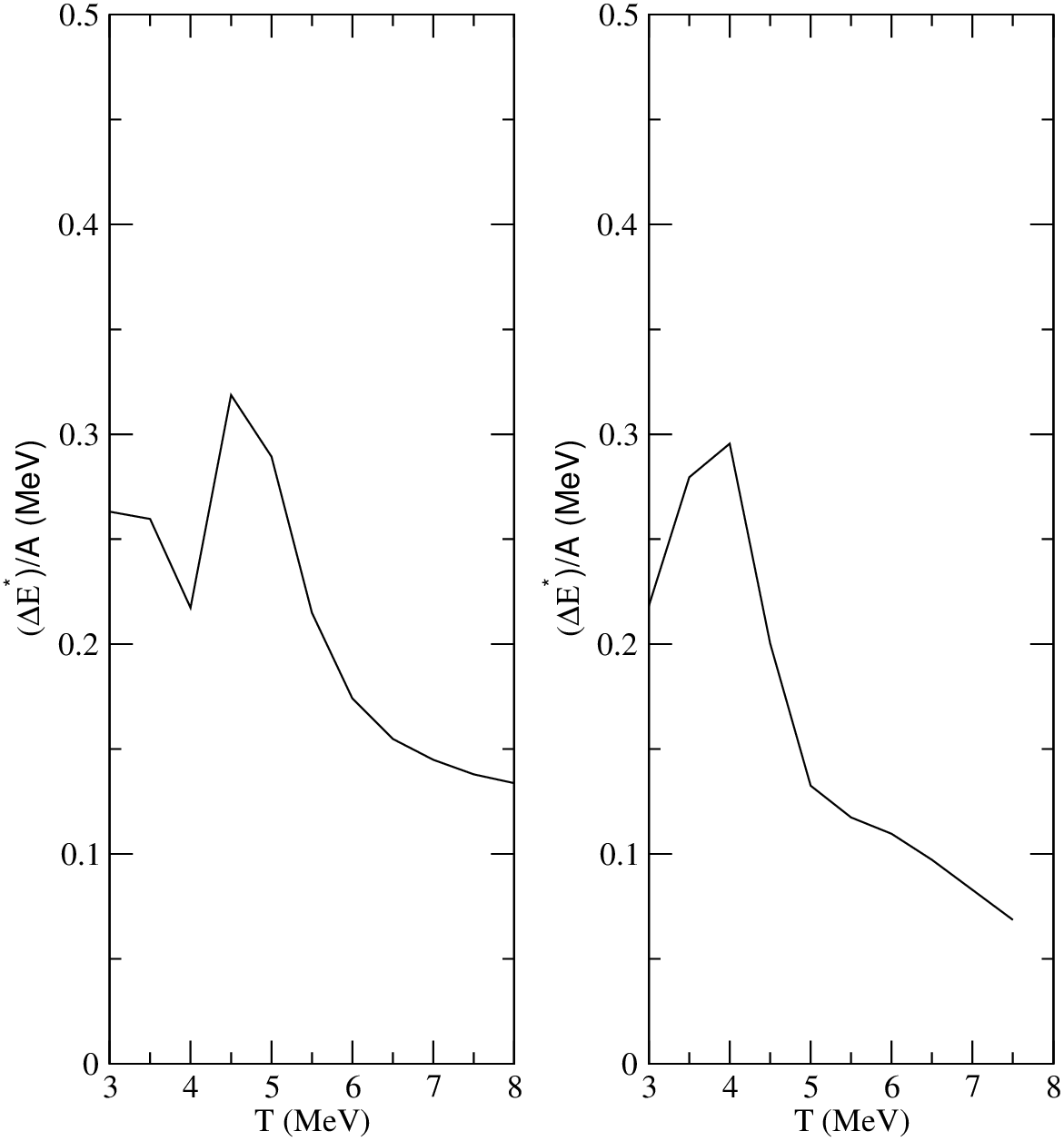}
\caption{ In the left panel we plot $E^*(146,64)/146-E^*(146,56)/146$
against $T$.  In the right panel we plot $E^*(146,64)/146-E^*(73,32)/73$
against $T$.  See section V for a discussion.  Only results using
CTM are shown; SMM gives very similar results.}
\end{figure}


\begin{thebibliography}{99}

\bibitem{Bondorf1} J. P. Bondorf, A. S. Botvina, A. S. Iljinov, I. N. 
Mishustin and K. Sneppen, Phys. Rep. {\bf 257}, 133 (1995).

\bibitem{Xi} H. Xi et al., Z. Phys. A {\bf 359}, 397 (1997).

\bibitem{Dagostino} M. D'Agostino et al., Phys. Lett. B {\bf 371}, 175 (1996).

\bibitem{Scharenberg} R.P. Scharenberg et al., Phys. Rev. C {\bf 64}, 
054602 (2001).

\bibitem{Das1} C. B. Das, S. Das Gupta, W. G. Lynch, A. Z. Mekjian and
M. B. Tsang, Phys. Rep. {\bf 406}, 1 (2005)

\bibitem{Tsang1} M. B. Tsang et al., Phys. Rev. C{\bf 64}, 054615 (2001)

\bibitem{Chaudhuri1} G. Chaudhuri, S. Das Gupta, W. G. Lynch, M. Mocko,
and M. B. Tsang, Phys. Rev. C{\bf 76}, 067601 (2007)

\bibitem{Jackson} A.S. Botvina, A.D. Jackson, I.N. Mishustin, 
Phys. Rev. E{\bf 62}, R64 (2000).

\bibitem{Botvina01} A.S. Botvina, I.N. Mishustin,
Phys. Rev. C {\bf 63}, 061601 (2001).

\bibitem{Mocko1} M. Mocko, Ph. D. Thesis, Michigan State University, 2006.


\end{thebibliography}
\end{document}